\documentclass[twocolumn]{aastex63}
\usepackage{amsmath,amstext,amssymb}
\usepackage{lipsum}  
\usepackage{xspace}
\usepackage{multirow}
\usepackage{appendix}
\usepackage[percent]{overpic} % in your preamble

\newcommand{\CMU}{McWilliams Center for Cosmology and Astrophysics, Department of Physics, Carnegie Mellon University, Pittsburgh, PA 15213, USA}

\newcommand{\backpop}{\texttt{BackPop}\xspace} 
\newcommand{\cosmic}{\texttt{COSMIC}\xspace}

\shortauthors{Maga\~na~Hernandez and Breivik}
\shorttitle{Lucky Strikes}
\begin{document}

\title{Lucky Strikes: On the Origins of GW190814 Through Isolated Binary Evolution}

\author[0000-0003-2362-0459]{Ignacio Maga\~na~Hernandez}
\affiliation{\CMU}
\email{imhernan@andrew.cmu.edu}

\author[0000-0001-5228-6598]{Katelyn Breivik}
\affiliation{\CMU}
\email{kbreivik@andrew.cmu.edu}

\begin{abstract}
The asymmetric nature of GW190814, particularly its mass ratio ($q \approx 1/10$), has made its astrophysical origin elusive. We explore isolated binary evolution as a potential explanation for GW190814’s formation. Using the binary population synthesis code \cosmic, and the \texttt{BackPop} sampling technique to map the observed parameters of GW190814 to the initial conditions of Zero Age Main Sequence binary stars while simultaneously inferring the astrophysical prescriptions for common envelope evolution, stable mass transfer and natal kick kinematics that are needed for its formation and eventual merger. We find that the initial conditions for the binary stellar population that forms GW190814 do not stand out significantly from massive star populations observed in the Local Group. Our \texttt{BackPop} simulations recover a dominant formation pathway where the first Roche overflow phase includes a common envelope evolution and the second Roche overflow phase remains stable. Our findings suggest that natal kicks imparted during compact object formation play the strongest role in forming GW190814-like systems. Specifically, our models require a low magnitude first natal kick (independent of direction) that prevents the binary from unbinding and a large second natal kick with its direction in the plane of the orbit and toward the binary's center of mass. The second natal kick strength and direction crucially increases the orbital eccentricity, leading to shorter delay times, and thus enabling mergers within a Hubble time. We estimate the chance probability for GW190814-like events that experience such a \emph{lucky} kick and find that it occurs in $\sim20\%$ of systems if natal kicks are randomly oriented. We discuss the astrophysical implications for the formation of asymmetric GW190814-like systems under the context of binary stellar evolution. 
\end{abstract}

 
\section{Introduction} \label{sec:Intro}

One of the most astrophysically interesting gravitational-wave (GW) detections reported by the LIGO/Virgo/KAGRA (LVK) Collaboration is GW190814---a compact binary coalescence (CBC) characterized by the most extreme mass ratio observed to date ($q \approx 1/10$)~\citep{Abbott2020}. This event involved a secondary compact object with a mass of approximately $2.6\ M_{\odot}$, placing it in a regime that challenges its classification: it may represent either the most massive known neutron star (NS) or the least massive black hole (BH) identified during the LVK’s third observing run (O3)~\citep{GWTC3-catalog}. Due to its atypical mass configuration, GW190814 is generally treated as an outlier~\citep{Essick:2021vlx} and excluded from standard binary black hole (BBH) population analyses as well as from neutron-star population constraints~\citep{GWTC3-pop,GWTC4-pop}. It is typically considered only in studies that analyze GW candidates independently of their astrophysical classification, for example across the entire CBC mass spectrum~\citep{GWTC4-pop}. Another challenge is that its lower-mass component falls within the presumed lower mass gap~\citep{Kalogera:1996ci,Bailyn:1997xt}. However, the recent event GW230529 had a primary mass of $2.5$--$4.5\ M_{\odot}$ with a confidently identified secondary NS component, bringing into question the very existence of this mass gap~\citep{massgap}. These events are similarly ambiguous, yet they differ significantly in mass ratio, since GW230529 only had $q \approx 0.4$~\citep{massgap}.  Even with the recent release of the Gravitational-Wave Transient Catalog 4.0 (GWTC-4.0)~\citep[e.g.,][]{GWTC4-catalog}, no other GW candidate has observed properties comparable to those of GW190814~\citep{GWTC4-pop}.

The low mass of the secondary object in GW190814, combined with its small localization volume, prompted extensive electromagnetic (EM) follow-up campaigns shortly after its discovery \citep{DES:2020ref,GravityCollective:2021kyg,deWet2021,Alexander:2021twj,Dobie:2021khu}. Despite deep and wide-field observations across multiple wavelengths, no EM counterparts were confidently detected. These non-detections disfavor the presence of a neutron star or other exotic compact object as the nature of the less massive component in the merger, however, that does not mean that a NS was not present but, that it was too faint to be detectable. For example, \citet{Zhu2021} studied the detectability for associated kilonovae signals from NS mergers during O3 \citep{Zhu2021}. While some studies have considered the possibility of the event occurring in an active galactic nucleus (AGN) environment where the lower mass CO is interpreted to be a NS \citep[e.g.][]{Tagawa+2021:2021ApJ...908..194T}, current constraints suggest this scenario is unlikely, though not entirely ruled out~\citep{Rowan:2024lla}. Consequently, GW190814 is most likely interpreted as an asymmetric BBH merger.

GW190814 has proven to be difficult to simulate with merger rates consistent with observations using isolated binary evolution models \citep[e.g.][]{Zevin2020,Broekgaarden+2021:2021MNRAS.508.5028B,Shao+2021:2021ApJ...920...81S,Tanikawa+2021:2021ApJ...910...30T,Olejak+2021:2021A&A...651A.100O,Ghodla+2022:2022MNRAS.511.1201G}.
Due to its asymmetric mass ratio, several works and have considered it's formation through dynamical assembly where hierarchical mergers may play a role \citep[][]{Gerosa+2021:2021NatAs...5..749G}. In particular, formation of GW190814-like systems in dense stellar clusters \citep{Flitter+2021:2021MNRAS.507..743F, Kritos+2021:2021PhRvD.104d3004K,Liu+2021:2021MNRAS.502.2049L,ArcaSedda+2021:2021ApJ...908L..38A,Ye+2024:2024ApJ...975...77Y} has been investigated extensively, though again, the predicted merger rates are too low to be consistent with observations. Other proposed hierarchical formation scenarios include formation in a triple system \citep{Lu+2021:2021MNRAS.500.1817L} or in the disk of an active galactic nucleus \citep{Tagawa+2021:2021ApJ...908..194T}.

In this paper, our goal is to assess whether GW190814-like systems can be reliably simulated and reproduced through isolated binary stellar evolution using the binary population synthesis library \cosmic\ \citep{Breivik2020}. To this end, we employ \backpop, in which binaries composed of zero-age main sequence (ZAMS) stars are evolved under a specified set of initial conditions and astrophysical prescriptions governing their subsequent evolutionary stages, ultimately leading to double compact object mergers. We then apply stochastic sampling methods to evaluate the consistency of the resulting populations with the observed parameters of GW190814. An advantage of \backpop\ is its ability to mitigate the curse of dimensionality that typically arises in other binary population synthesis simulations which rely on large grids over both initial conditions and astrophysical prescription parameter space.

Previous work by \citet{Andrews2020} characterized the initial-condition parameter space for the first gravitational-wave event, GW150914~\citep{LIGOScientific:2016aoc}. Building on this, \citet{Wong2023} expanded the analysis by sampling over the astrophysical assumptions governing binary evolution. We further extend the \backpop\ framework by introducing greater flexibility. Specifically, we make no assumptions regarding the universality of mass transfer events during binary evolution. For example, if a system undergoes two common-envelope phases, the ejection efficiencies for each stage are treated as independent parameters controlling the relevant astrophysics. The same principle is applied to stable mass transfer stages where the accretion efficiency for each star can be any fixed value between $0$ and $1$. In addition, we explicitly track the natal kicks associated with each compact object formation, keeping track of both their magnitude and direction.

In this paper, we focus on the progenitor analysis of GW190814, an outlier of the binary black hole population as described above. While this event has several proposed formation scenarios relying on dynamical interactions, here we aim to answer the question: what astrophysical prescriptions is isolated binary evolution are required to form GW190814-like mergers?

The structure of this paper is as follows. In Section~\ref{sec:Methods}, we introduce the statistical framework underlying \backpop. Section~\ref{sec:Results} presents our inference results. In Section~\ref{sec:Astro}, we explore the astrophysical implications of our findings, and in Section~\ref{sec:Conclusion}, we outline future directions and prospects. Throughout this work, we adopt a flat $\Lambda$CDM cosmology with parameters from the Planck 2015 results: a Hubble constant $H_0 = 67.8,\mathrm{km,s^{-1},Mpc^{-1}}$ and a matter density parameter $\Omega_m = 0.308$ \citep{Planck2015}.

\section{Methods} \label{sec:Methods}
\subsection{Statistical Framework}
We define the progenitor parameters of binary stars at ZAMS by $\theta=\{ m_1,m_2,\log_{10}t,\log_{10}Z \}$ where \( m_1 \) and \( m_2 \) are the stellar masses, \( t \) is the orbital period, and \( Z \) is the metallicity. We explicitly fix the ZAMS eccentricity to be zero.  

The binary stellar evolution is governed by a set of hyperparameters $\Lambda=\{ \alpha_1,\alpha_2,f_{\rm{lim},1},f_{\rm{lim},2} \}$ which control the physical prescriptions for mass transfer events. Specifically, we implement independent common envelope efficiency parameters \( \alpha_1 \) and \( \alpha_2 \), and stable mass transfer (SMT) accretion limits \( f_{\rm{lim},1} \) and \( f_{\rm{lim},2} \), each representing a distinct evolutionary pathway.

We note that in all common envelope events, we calculate the binding energy parameter, $\lambda_{\rm{CE}}$, directly based on the properties of the donor star following the fits described in Appendix A of \citet{Claeys+2014:2014A&A...563A..83C} assuming that no internal energy is used to unbind the envelope. Since both $\alpha$ and $\lambda_{\rm{CE}}$ are used in the common envelope calculation, we report the combined parameter $\alpha\lambda_{\rm{CE}}$ in our results for binaries which experience a common envelope in their evolution. To keep our setup as simple as possible, we fix the mass transfer stability criteria for stars crossing the Hertzsprung Gap using the critical donor–accretor mass ratio, $q_{\rm{HG}}=3.0$, that determines the boundary between stable Roche‑lobe mass transfer and the onset of a common envelope.

We also track the supernova natal kick properties for each core collapse event and subsequent compact object formation. These are denoted by $X = \{v_i, \theta_i, \phi_i, \omega_i \}$ for $i=1,2$ where \( v_i \) is the kick velocity magnitude, \( \theta_i, \phi_i \) are the kick angles and  $\omega_i$ is the phase of the orbit when the compact object forms. These parameters are critical, as they influence whether the binary remains bound or merges within a Hubble time. We note that a mass-loss kick \citep{Blaauw+1961:1961BAN....15..265B} is applied to all systems which lose mass during the core collapse where we assign remnant masses following the `delayed' prescription of \citet{Fryer+2012:2012ApJ...749...91F}. Since this prescription does not differentiate the treatment of neutron stars or BHs separately, we artificially reduce the maximum neutron star mass to $1\,M_{\odot}$ for ease in the sampling process.  
All together, the initial binary conditions, binary interaction hyperparameters, and natal kicks correspond to a 17-dimensional parameter space.

We denote the GW observables by $\theta_{\rm{GW}}=\{m_{1,\rm{GW}},m_{2,\rm{GW}}\}$ representing the source-frame masses of the primary and secondary compact objects in the merger. The binary population synthesis code \cosmic\ provides a deterministic mapping from the initial binary conditions, hyperparameters, and SN kicks to the predicted merger masses,
\begin{equation}\label{eqn:cosmic}
    \theta_{\rm{GW}} = f(\theta,\Lambda,X).
\end{equation}

Using this model, we write the posterior distribution over the initial conditions, hyperparameters, and natal kicks as,
\begin{equation}\label{eqn:posterior1}
    p(\theta,\Lambda,X|x_{\rm{GW}}) \propto \int \mathcal{L}(x_{\rm{GW}}|\theta_{\rm{GW}})p(\theta_{\rm{GW}}|\theta,\Lambda,X) d\theta_{\rm{GW}}
\end{equation}
where $\mathcal{L}(x_{\rm{GW}}|\theta_{\rm{GW}})$ is the individual event GW likelihood function and $p(\theta_{\rm{GW}}|\theta,\Lambda,X)$ is the population model linking individual GW parameters in terms of binary evolution hyperparameters and corresponding progenitor stars initial conditions. Recognizing this, we make use of the deterministic model defined in Equation~\ref{eqn:cosmic} and write, $p(\theta_{\rm{GW}}|\theta,\Lambda,X) = \delta(\theta_{\rm{GW}} - f(\theta,\Lambda,X))$ to perform the integration in Equation 2 to obtain,
\begin{equation}\label{eqn:posterior2}
    p(\theta,\Lambda,X|x_{\rm{GW}}) \propto \mathcal{L}(x_{\rm{GW}}|f(\theta,\Lambda,X))p(\theta,\Lambda,X)
\end{equation}
where we have explicitly written the prior on model hyperparameters and initial conditions $p(\theta,\Lambda,X)$. 

\begin{figure*}
\centering
\begin{overpic}[width=\textwidth]{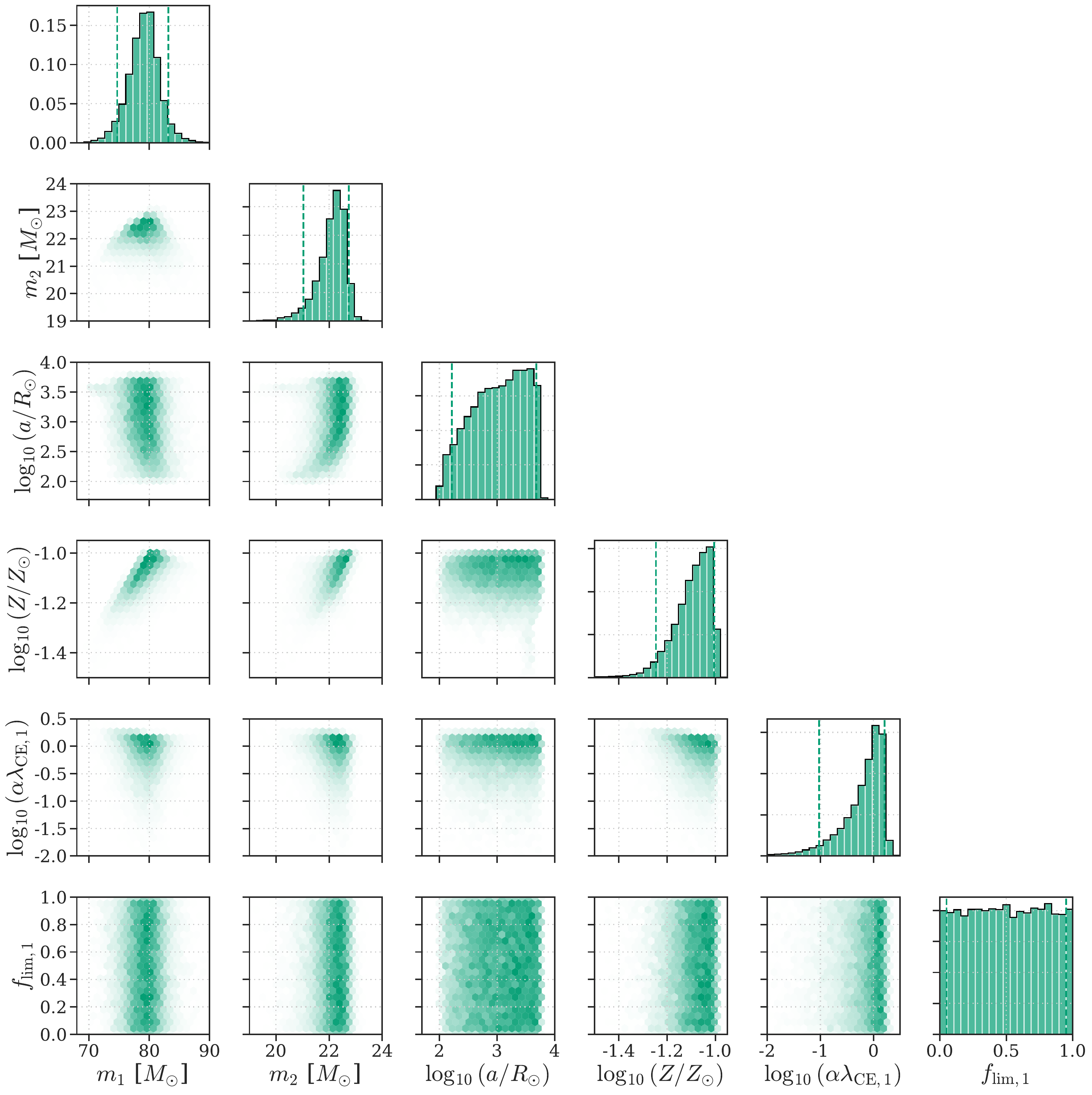}
    % Place inset at (x,y) in percent of main figure width/height
    \put(60,60){ % adjust these numbers to position inset
        \includegraphics[width=0.4\textwidth]{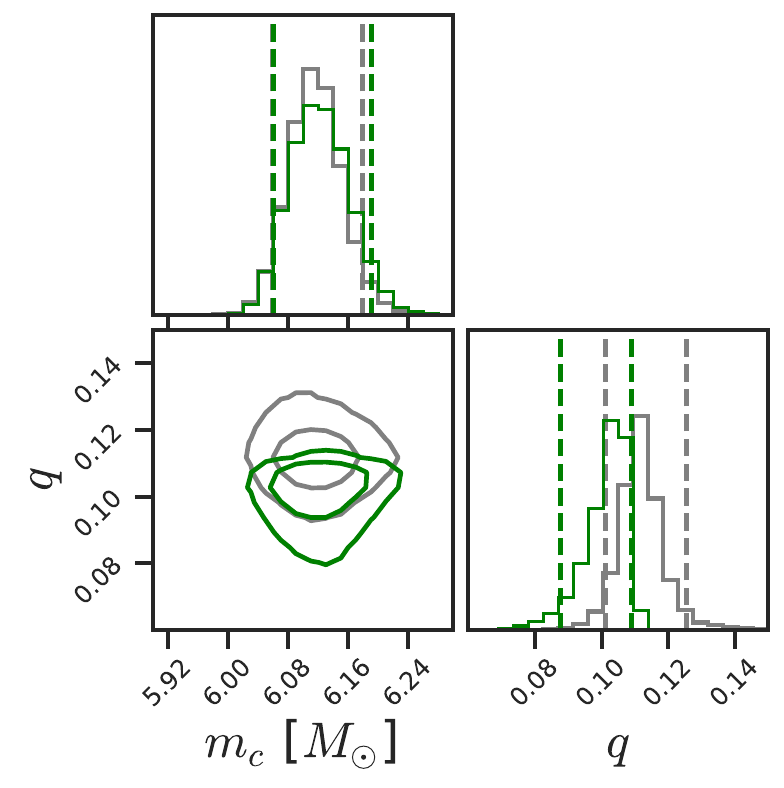}
    }
\end{overpic}
\caption{Corner plot showing the initial conditions for the binary stars at ZAMS that are consistent with the formation of the BBH merger GW190814. We also show the binary evolution hyperparameters $\alpha\lambda_{\rm{CE},1}$ and $f_{\rm{lim},1}$, demonstrating that producing a GW190814-like event requires finely tuned common envelope efficiency and mass transfer limits. In the top-right panel, we show the inferred GW190814 source frame chirp mass ($m_c$) and mass ratio ($q$) after evolving our initial conditions and hyperparameters. We overplot the corresponding LVK mass measurements \citep{Abbott2020} and find broad consistency, though we note that our simulations tend to prefer slightly more assymetric mass ratios.}
\label{fig:corner}
\end{figure*}

\subsection{Inference Setup}

To sample the posterior in Equation~\ref{eqn:posterior2}, we approximate the likelihood using publicly available posterior samples for GW190814. We perform density estimation on the marginalized posterior distribution over the source-frame primary and secondary masses. This enables direct evaluation of the binary population synthesis output within the likelihood function, as defined by Equation~\ref{eqn:cosmic}, and required for inference in Equation~\ref{eqn:posterior2}. For technical details on constructing the source-frame mass likelihood from GW posterior samples, we refer the reader to Appendix~\ref{appendix:reweighting}.

We simulate the formation of GW190814 using the \texttt{BackPop} framework in combination with \cosmic, a binary population synthesis code designed to model the evolution of compact object binaries from zero-age main sequence to merger \citep{Breivik2020}. For this study, we used \cosmic v3.6.1 which applies the kick model described in \citet{Wagg+2025:2025OJAp....8E..85W} that is based on \citet{Pfahl+2002:2002ApJ...573..283P}. We sample the full parameter space described in Section~\ref{sec:Methods}, including the initial binary conditions at ZAMS, the astrophysical hyperparameters governing binary evolution, and the core-collapse natal kick magnitudes and directions imparted during compact object formation. We take uniform priors on all parameters within physically relevant ranges and within the validity of \cosmic, these are shown in Table~\ref{tab:priors}.

\begin{deluxetable}{cc}
\tablecaption{All parameters are assigned uniform priors within the ranges specified. \label{tab:priors}}
\tablehead{
\colhead{Parameter} & \colhead{Range}
}
\startdata
$m_{1}\,[M_{\odot}]$        & $[50,\,150]$ \\
$q$            & $[0.01,\,1]$ \\
$\log_{10}(t_{\mathrm{b}}/\,\mathrm{days})$ & $[0,\,\log_{10}(5000)]$ \\
$\log Z$       & $[-4,\,\log_{10}(0.03)]$ \\
\hline
$\alpha_{1}$   & $[0.1,\,20]$ \\
$\alpha_{2}$   & $[0.1,\,20]$ \\
$f_{\rm{lim,1}}$ & $[0,\,1]$ \\
$f_{\rm{lim,2}}$ & $[0,\,1]$ \\
\hline
$v_{k1}\,[\mathrm{km\,s^{-1}}]$ & $[0,\,500]$ \\
$\theta_{1}\,[\rm{deg}]$   & $[0,\,360]$ \\
$\phi_{1}\,[\rm{deg}]$     & $[-90,\,90]$ \\
$\omega_{1}\,[\rm{deg}]$   & $[0,\,360]$ \\
$v_{k2}\,[\mathrm{km\,s^{-1}}]$       & $[0,\,500]$ \\
$\theta_{2}\,[\rm{deg}]$   & $[0,\,360]$ \\
$\phi_{2}\,[\rm{deg}]$     & $[-90,\,90]$ \\
$\omega_{2}\,[\rm{deg}]$   & $[0,\,360]$ \\
\enddata
\end{deluxetable}

As noted by \citet{Wong2023}, systems like GW190814 pose significant challenges for sampling due to their rarity and sensitivity to specific evolutionary pathways. Indeed, even in the case for a typical merging system like GW150914, \citet{Wong2023} performed a forward-modeling initialization step to choose regions of the initial binary parameter space that preferentially lead to BBH mergers before performing the Markov Chain Monte Carlo sampling step. In this work, we explore the use of nested sampling with a specific goal of avoiding this initialization step. This choice naturally leads to other benefits of nested sampling including the explicit calculation of the evidence integral which could be particularly useful in future studies which hierarchically model systems using \backpop. 

To perform inference, we use the nested sampling library \texttt{Nautilus} \citep{Lange:2023ydq}, which efficiently explores the posterior distribution over initial conditions, hyperparameters, and natal kicks as defined in Equation~\ref{eqn:posterior2}. \texttt{Nautilus} leverages neural networks to boost sampling efficiency, which is particularly important for systems like GW190814. In such cases, most prior draws result in binaries that do not merge within a Hubble time, leading to large regions of flat likelihood—sampling plateaus that are difficult to navigate with traditional methods. We use the following sampler settings: we set the number of live points, \texttt{nlive=3000} to ensure thorough exploration of the posterior landscape. To guarantee convergence, we require an effective sample size of at least $\sim1000$ per dimension. For this work, we proceed with a conservative estimate and set \texttt{neff=30000}. All other parameters are left at their default values.

\section{Results} \label{sec:Results}
Since the \backpop\ sampling procedure maps specific evolutionary channels, there are significant portions of the parameter space that are completely uncorrelated. For example, if a BBH merger forms through two mass transfer phases (one each from each stellar component) and both of these mass transfer phases remain stable throughout the evolution, assumptions for the common envelope efficiency are completely moot. Restricting our attention to only the set of parameters are correlated thus effectively follows the preferred formation channel that is mapped by the nested sampling. For the sake of presentation, we do not plot the entire 17-dimensional parameter space from our simulations in the main text; a corner plot of the full posterior sample is shown in Figure~\ref{fig:fat-corner} in Appendix~\ref{appendix:full_posterior}.

Figure~\ref{fig:corner} presents the results of our simulation, highlighting the binary interaction parameters—specifically the common envelope efficiency and stable mass transfer accretion efficiency- alongside the initial binary conditions required to produce a GW merger consistent with GW190814. As a consistency check, the inset in Figure~\ref{fig:corner} shows the chirp mass and mass ratio samples inferred by \texttt{BackPop} when we forward evolve them via \cosmic, which are broadly consistent with the source-frame chirp mass ($m_c$) and mass ratio ($q$) reported for GW190814 \citep{Abbott2020}. 

\begin{figure*}[!t]
\centering
\includegraphics[width=\textwidth]{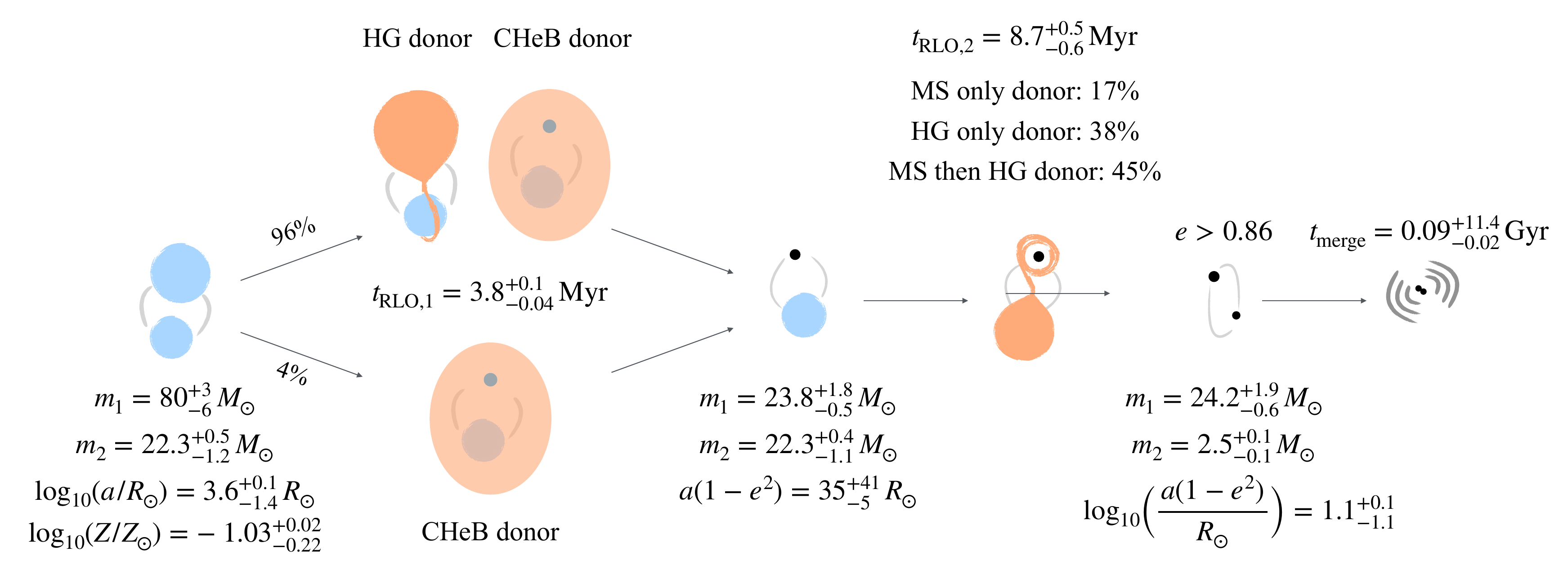}
\caption{Schematic for the formation pathway of GW190814-like binaries using \cosmic. The first RLOF is dominated by common envelope evolution that occurs during the initially more massive star's core helium burning phase (CHeB) phase following a stable mass transfer initiated during the Hertzsprung Gap (HG). The second RLOF is stable with Eddington limited accretion onto the first-formed BH. A strong natal kick for the second-formed compact object drives the eccentricity that allows the binary to merge within a Hubble time. The peak and $90\%$ intervals of the distributions for the binary properties and event times are shown along with the fractions of different stellar evolutionary stages at which the population that undergoes each RLOF.}
\label{fig:schematic}
\end{figure*}

The preferred ZAMS masses range from $\sim75-85\,M_{\odot}$ and $\sim20-23\,M_{\odot}$ with a strong preference for mass ratios of $q\simeq0.29$. While we sample ZAMS orbital periods uniformly in $\log_{10}(P_{\rm{orb}}/\rm{day})$, we plot the semimajor axis as $\log_{10}(a/R_{\odot})$ here since it can be easily compared to the radius evolution of the primary star. We find a wide range of initial semimajor axes and a slight preference toward the widest semimajor axis (near $5800\,R_{\odot}$) at which the primary would fill its Roche lobe while crossing the Hertzsprung Gap.
The metallicity distribution peaks strongly near $Z=0.1\,Z_{\odot}$ with a tail that extends to $0.03\,Z_{\odot}$.

The ZAMS masses are highly correlated with metallicity, with higher masses being initiated with higher metallicities. This is directly due to \cosmic's wind mass loss prescription which follows \citet{Vink+2001:2001A&A...369..574V} and \citet{Vink+2005:2005A&A...442..587V} for main sequence and Wolf-Rayet stars respectively. This prescription applies a metallicity-specific mass loss rate that increases with increasing metallicity. The ZAMS semimajor axis and secondary mass are also correlated such that lower secondary masses are matched with smaller semimajor axes. This directly facilitates the first Roche-lobe overflow (RLOF) occurring when the primary mass is crossing the Hertzprung Gap.

We find an obvious correlation between the combined common envelope ejection efficiency and envelope binding parameter for the first RLOF ($\alpha\lambda_{\rm{CE,1}}$) and the ZAMS properties of the binary. This is especially apparent where initially wider orbits have a larger potential orbital energy reservoir that can be used to unbind the envelope of the primary donor star. Similarly, we find a slight correlation between $\alpha\lambda_{\rm{CE,1}}$ and the ZAMS masses since higher masses contribute to larger orbital energies at RLOF. This is further compounded with higher masses leading to larger radii such that the RLOF occurs in wider orbits\footnote{We note that the radius of the stars in the fitting formulae used by \cosmic may be over predicted \citep[e.g.][]{Agrawal+2020:2020MNRAS.497.4549A}. A decrease in the radii of the stars would lead to preferences toward larger $\alpha\lambda_{\rm{CE,1}}$.}. Finally, we find an anticorrelation between $\alpha\lambda_{\rm{CE,1}}$ and metallicity that is due to the effect of higher wind mass loss rates widening the binary orbit for stars with higher metallicities. All together the correlations between initial binary properties and common envelope ejection efficiency suggest a general preference for $\alpha\lambda_{\rm{CE,1}}\sim2$ where larger $\alpha_1$ tracks lower masses, shorter initial periods, and lower metallicities.

We also find a broad, uncorrelated distribution of possible stable mass transfer accretion efficiencies for the second RLOF ($f_{\rm{lim},1}$). This is due to the accretion efficiency of stable mass transfer being limited to $10$ times the Eddington accretion rate of the BH, thus removing any dependence on $f_{\rm{lim},1}$. We also do not find any correlations with the common envelope ejection efficiency during the second RLOF ($\alpha_2$). Upon manual inspection, all binaries in our sample do not experience a common envelope evolution during the second RLOF. It is thus impossible to calculate the combined $\alpha\lambda_{\rm{CE,2}}$ parameter in this case. Finally, we find no correlations between $f_{\rm{lim},2}$ and any other initial binary parameters, hyperparameters, or natal kicks. Together, this set of correlations with $\alpha\lambda_{\rm{CE,1}}\sim2$, and lack thereof for $\alpha_2$, $f_{\rm{lim},1}$, and $f_{\rm{lim},1}$ describe an evolutionary history for GW190814-like systems that undergo a common envelope evolution during the first RLOF and a stable, but nonconservative, mass transfer episode during the the second RLOF. 

A summary of the formation history of our simulated GW190814-like systems is shown in Figure~\ref{fig:schematic}. The vast majority of the simulated binaries undergo a very short period of stable mass transfer in the first RLOF initiated by the initially more massive star while it is traversing the Hertzsprung Gap (HG in the figure). Once the star begins core helium burning (CHeB in the figure), a common envelope is unavoidable since the radius of the initially more massive star puffs up to completely envelope its stellar companion. Following the common envelope, the initially more massive star eventually reaches core collapse and produces the more massive BH with a low kick that is characteristic for BHs with masses above $20\,M_{\odot}$ (see Figure~\ref{fig:fat-corner} in Appendix B). The second RLOF is always stable due to the near-equal mass ratio of the BH and secondary star, but Eddington limited such that the first-formed BH's mass increases only by $\sim0.4\,M_{\odot}$. Finally, the second compact object forms with a strong kick that excites eccentricity to the newly formed double compact object binary. We find that the formation eccentricity distribution of the double compact object a \emph{minimum} of $e>0.86$ with the distribution peaking at $e\sim0.99$. The minimum eccentricity is directly tied to our requirement that the binary merges within a Hubble time.

\begin{figure}
\centering
\includegraphics[width=0.48\textwidth]{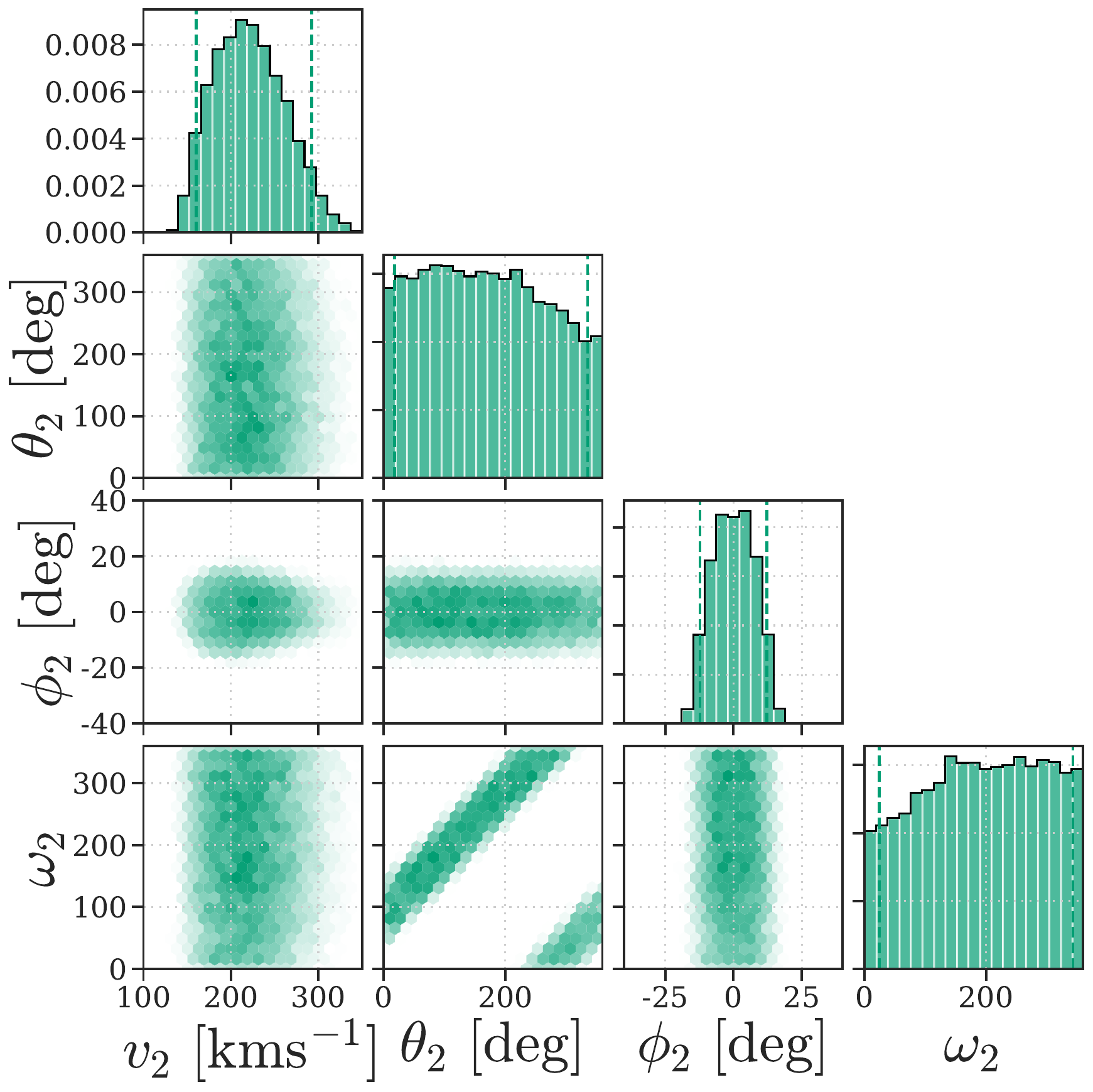}
\caption{Corner plot showing the marginalized posterior distribution over the SN natal kick parameters for the second-formed compact object. We display the kick magnitude $v_2$, the kick angles $\theta_2$ and $\phi_2$, and the phase of the orbit when the compact object forms $\omega_2$. Our results indicate that the system requires a \emph{lucky strike}—a kick with $v_2 \gtrsim 150\,\mathrm{km\,s^{-1}}$ in the plane of the orbit and perpendicular to the direction of motion. This configuration leads to a double compact object binary with a large eccentricity that rapidly circularizes and shrinks due to GW emission, thus enabling a merger within a Hubble time.}
\label{fig:kicks}
\end{figure}

Figure~\ref{fig:kicks} shows the posterior distributions for the supernova natal kick parameters associated with the formation of the second-formed compact object. Our results indicate that, for the binary to remain bound and eventually merge, a large natal kick velocity is required —specifically, one that is in the plane of the orbit ($\phi_2$ centered about $0\,\rm{deg}$) and aligned perpendicular to the direction of motion such that the azimuthal kick angle, $\theta_2$, trails the compact object progenitor's orbital phase, $\omega_2$ by $90\,\rm{deg}$. We find that the minimum natal kick velocity that produces a merger is $v_2>128\rm{\,km\,s^{-1}}$ with strongly kicks being slightly more tightly constrained to the plane of the orbit. Kick angles which extend slightly above or below the orbital plane and away from the orbital motion of the compact object progenitor are allowed, but only in a small window of correlated angles. This configuration represents a \emph{lucky strike}, where the kick enhances the system's eccentricity without disrupting it. 

To quantify the likelihood of such a favorable kick, we compute the posterior probability over the solid angle of the kick direction and compare it to the isotropic case. We find that the probability of a kick occurring within the required directional solid angle (estimated to have an area $\delta \Omega \approx 2.57 \ \rm{sr}$) is approximately $P_{\rm{kick}} = \delta \Omega/4\pi \approx 0.20$. This low probability suggests that the formation rate of GW190814-like systems is intrinsically rare even though the ZAMS binary properties do not occupy a low-probability region of parameter space. If the formation pathway we propose is indeed responsible for GW190814, then the event rate should scale proportionally with $P_{\rm{kick}}$. A full rate calculation, however, requires integrating over the metallicity-dependent star formation history and accounting for binary survival across cosmic time—a task we leave for future work.

Since black holes with masses in excess of $20\,M_{\odot}$ are generally expected to form with negligible kicks both from a theoretical \citep[e.g.][]{Fryer+2022:2022ApJ...931...94F, Janka+2024:2024Ap&SS.369...80J, Burrows+2025:2025ApJ...987..164B} and observational perspective \citep[e.g.][]{Balbinot+2024:2024A&A...687L...3B, MarinPina+2024:2024A&A...688L...2M}, we investigate how our sampling process changes when restricting natal kicks for the first formed black hole to be zero. This removes the $v_{k1}$, $\theta_1$, $\phi_1$, and $\omega_1$ natal kick parameters from our simulations, though we note that mass loss kicks can still alter the orbital period and eccentricity of the binary following the formation of the first black hole. A corner plot of the full posterior sample is shown in Figure~\ref{fig:fat-corner_fixed} in Appendix~\ref{appendix:full_posterior}. 

\begin{figure}
\centering
\includegraphics[width=0.48\textwidth]{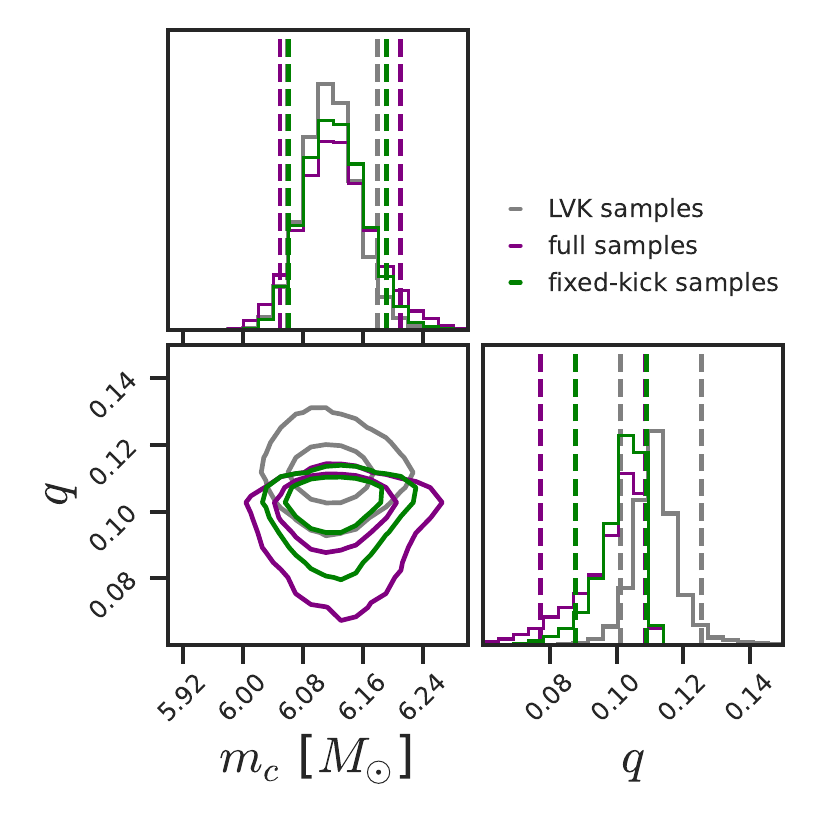}
\caption{We show the simulated distributions for chirp mass and mass ratio of merging BBHs for the full 17-parameter simulation in green and the fixed-kick 13-parameter simulation in purple. As in Figure~\ref{fig:corner}, we also show the LVK posteriors for the chirp mass and mass ratio for comparison in grey. Reducing the parameter space by fixing the natal kick of the first-formed black hole to be zero slightly widens the chirp mass -- mass ratio distribution, primarily extending it to lower mass ratios.}
\label{fig:compare}
\end{figure}

Fixing the natal kicks to zero for the first-formed black hole does not qualitatively change our results, however, there are slight quantitative changes that arise. These are most clearly seen in comparisons between the chirp mass and mass ratio of the BBH merger as shown in Figure~\ref{fig:compare} where the LVK parameter estimation posteriors are shown in grey, the distribution from the full 17-parameter simulation is shown in green, and the distribution from the fixed natal kick 13-parameter simulation is shown in purple. Fixing the natal kick for the first-formed black hole produces merging BBHs with slightly wider posterior distributions for the chirp mass and mass ratio primarily extending the mass ratio distribution even farther below the LVK contours. 

This extension reflects the more rigid bounds of the allowed evolution for the GW190814-like progenitors. In our fixed-kick simulations, these slight changes primarily influence the binary semilatus rectum after the formation of the first black hole which has a cascading effects primarily for the progenitor evolution from ZAMS to the formation of the first black hole. When fixed-kicks are assumed for the first-formed black hole, we find that the primary ZAMS mass distribution tightens near $80\,M_{\odot}$ such that the fraction of stars experiencing a stable mass transfer when the primary is crossing the Hertzsprung Gap increases from $94\%$ to $98\%$. The evolution of each binary once the first black hole is formed is largely unchanged from the statistics reported in the schematic in Figure~\ref{fig:schematic} save for slightly narrowed distributions around the peak of the two black hole mass distributions which extend the mass ratio toward lower values. Given the strong agreement between the 17-parameter and 13-parameter simulations, we argue that our results are statistically robust.

\section{Astrophysical Interpretation, Discussion, and Caveats} \label{sec:Astro}

Having recovered a formation history for GW190814-like binaries simulated by \cosmic\ with \texttt{BackPop}, we now consider how our results fit into a wider astrophysical context. First we consider the relative likelihood of the preferred ZAMS properties and composition of the binary. Then we discuss the strong preference toward common envelope that is initiated due to stellar evolution during an initially stable mass transfer episode. We finish with a discussion on the rarity of the preferred strength and orientation of the second natal kick. 

The initial masses (and therefore mass ratio) and semimajor axes of our GW190814-like progenitors are not particularly exotic. When comparing to the BBH merger mass ratio of $q_{\rm{BBH}}\sim0.15$, a ZAMS mass ratio of $q_{\rm{ZAMS}}\sim0.3$ is decidedly not extreme. The individual mass distributions themselves are broadly consistent with observations of nearby massive O stars, though at $Z\simeq0.1\,Z_{\odot}$ no stars with $M\simeq80\,M_{\odot}$ are known in the Local Group \citep{Camacho+2016:2016A&A...585A..82C}. That said, the Small Magellanic Cloud hosts several massive stars with masses $M\gtrsim80\,M+{\odot}$ and metallicities of $Z\simeq0.2\,Z_{\odot}$ \citep{Massey+2002:2002ApJS..141...81M}. Finally, the preference for large ZAMS semimajor axes with $a\sim4000\,R_{\odot}$ and our enforcement of the ZAMS eccentricity to be $e=0$ is qualitatively consistent with observations of massive stars \citep[][]{Sana+2012:2012Sci...337..444S}. We note, however, that enforcing circular binaries leads to a shift in toward lower semimajor axes on average since the first RLOF is expected to occur near the semilatus rectum such that larger ZAMS eccentricities can lead to RLOF in binaries with larger ZAMS semimajor axes. 

A key feature in the formation history of our simulated GW190814-like binaries is the dominance of a transition during the first RLOF from stable mass transfer to a common envelope evolution mediated by the primary star evolving to begin core helium burning during the RLOF. This requirement for a common envelope evolution which facilitates a significant orbital shrinking phase during the first RLOF casts fairly significant uncertainty on our simulations, because \cosmic\ does not account for how a donor star responds to RLOF during stable mass transfer. Further, as described above in Section~\ref{sec:Methods}, we rely on the fits to the \citet{Pols+1998:1998MNRAS.298..525P} stellar evolution grid described in Appendix A of \citet{Claeys+2014:2014A&A...563A..83C} to calculate the common envelope binding energy parameter $\lambda_{\rm{CE}}$. Our calculated values for $\lambda_{\rm{CE}}$ depend on the properties of the donor at RLOF and vary between $0.069\leq\lambda_{\rm{CE}}\leq0.217$, where lower $\lambda_{\rm{CE}}$ pairs with a larger $\alpha_{\rm{CE}}$ such that the binary doesn't shrink too much during the first RLOF. More recent studies \citep[e.g.][]{Klencki+2021:2021A&A...645A..54K} suggest that these $\lambda_{\rm{CE}}$ values may be too large and might lead to a merger during the first common envelope. 

An alternative scenario is that the mass transfer remains stable throughout the entire first RLOF. Indeed, studies which simulate binary stars with detailed stellar evolution find that mass transfer can remain stable in a wider range of orbital configurations than traditionally assumed in rapid binary population synthesis codes like \cosmic\ \citep[e.g.][]{Gallegos-Garcia+2021:2021ApJ...922..110G}. Since our formation scenario prefers that the secondary mass does not gain significant mass during the first mass transfer, it could be possible that the binary begins with a smaller ZAMS semimajor axis and undergoes strongly non-conservative stable mass transfer \citep[e.g.][]{Olejak+2025:2025arXiv251110728O}. Since \cosmic\ does not perform detailed stellar evolution calculations during RLOF, we leave a detailed study of the formation of GW190814-like systems for future work. We note, however, that our posterior samples provide an excellent starting target for such simulations. 

The most physically robust predictions from our simulations is the requirement of a strong natal kick oriented in the plane of the orbit and toward the center of mass of the binary, perpendicular to the direction of motion of the second-formed compact object. This finding agrees strongly with the results of \citet{Oh2023}, though perhaps not unsurprisingly since they also use \cosmic\ in their simulations. Studies which perform $>1$D simulations of core collapse also find that very massive neutron stars or low mass black holes (i.e. $\sim2\leq M_{\rm{CO}}/M_{\odot}\leq5$) likely form with strong natal kicks \citep[e.g.][]{Janka+2024:2024Ap&SS.369...80J,Burrows+2025:2025ApJ...987..164B}. However, we note that our preferred kick strength of $v_{k,2}\simeq220\,\rm{km\,s^{-1}}$ is generally lower than the $\sim300-1000\rm{km\,s^{-1}}$ kicks produced in some simulations. A potential route toward observational confirmation of large kicks being imparted to compact objects in this mass range is the relative rate of dormant compact object binaries discovered by Gaia and GW mergers detected by ground-based detectors \citep{Fishbach+2025:2025arXiv250808986F}.

\section{Conclusions and Future Work}\label{sec:Conclusion}

In this work, we have investigated the formation of GW190814 through isolated binary stellar evolution using  the \cosmic\ population synthesis code in combination with the \backpop\ sampling framework. Our analysis shows that GW190814-like system can be produced without requiring progenitor binary stars that are exceptionally rare compared to observed stellar populations in the Local Group. Instead, we find that what allows for GW190814-like systems to merge is due to 1) the sequence of binary interactions and 2) the natal kick imparted during the formation of the second compact object.

We find that the dominant formation channel involves a first RLOF phase that proceeds through CE evolution. This is then followed by a second RLOF that remains stable and non-conservative. The formation of the first compact object requires a low-magnitude natal kick to ensure that they binary system remains bound, while (most importantly) the second CO requires a large natal kick at formation, oriented along the orbital plane and towards the binaries center of mass. This \emph{lucky strike}, increases the orbital eccentricity ($e \gtrsim 0.86$), enabling the system to rapidly circularize and shrink due to GW emission, allowing GW190814-like systems to merge within a Hubble time. Fixing the natal kick to zero in the formation of the first compact object does not change this result and only slightly widens the distributions for the chirp mass and mass ratio of our simulated BBH mergers.

Although the progenitor binary stars are not unusual, the probability of such a favorable natal kick geometry is intrinsically low. We estimate that only $\sim 20\%$ of systems experience the required kick
configuration compared to isotropic natal kick distributions. This suggests, that the formation and sequential merger of GW190814-like events is potentially driven by natal kick dynamics rather than fine-tuned binary stellar initial conditions or physical prescriptions--owing to the observed rarity of such events.

Looking ahead, the analysis presented in this paper can be expanded to a broader set of gravitational-wave events. By applying our inference framework to other GW events, we can begin to map how metallicity, mass transfer efficiency, and natal kick distributions vary across the population. This will allow us to test whether the \emph{lucky strike} configuration inferred for GW190814 is unlikely or part of a broader distribution of outcomes. By performing population inference across GW catalogs, we can potentially place constrains on stellar binary physics prescriptions as population-level predictions that can be directly compared with the observed merger rates and population distribution predictions from GW observations.

Rare events like GW190814 provide a unique opportunity to leverage on the physics of isolated binary evolution because they probe the tails of the distribution where standard assumptions can break down. By systematically confronting models with both common and rare mergers, we can identify the physical processes that most strongly shape the diversity of binary stars that can give rise to these mergers. As gravitational-wave catalogs continue to expand, we expect that this strategy will enable constraints on the astrophysics of binary stellar evolution.

\section{Acknowledgements}
The authors would like to thank Antonella Palmese, Brendan O'Connor and Yeajin Kim for useful discussions. IMH is supported by a McWilliams postdoctoral fellowship at Carnegie Mellon University. KB is supported by NSF Grant AST-2510583. This research has made use of data or software obtained from the Gravitational Wave Open Science Center (gwosc.org), a service of the LIGO Scientific Collaboration, the Virgo Collaboration, and KAGRA.

\appendix
\onecolumngrid
\section{Likelihood Construction from Density Estimation} \label{appendix:reweighting}

The LVK collaboration releases posterior samples for individual gravitational-wave events, including GW190814. These samples characterize the detector-frame component masses, luminosity distance, and other parameters inferred from the observed signal. The initial parameter estimation is performed using a default prior that is uniform in detector-frame masses and proportional to $D_L^2$ for the luminosity distance \citep{Abbott2020}.

To perform hierarchical inference, we must estimate the likelihood function from these posterior samples. This is equivalent to recovering the posterior distribution in the source frame under uniform priors on all parameters. Transforming to the source frame requires assuming a fiducial cosmological model (e.g., Planck 2015 \citep{Planck2015}), which introduces a Jacobian factor that modifies the original prior. Specifically, the default PE prior transforms to:
\[
\pi_{\text{PE}}(m_{1,\text{src}}, m_{2,\text{src}}, z) \propto \pi_{\text{PE}}(m_{1,\text{det}}, m_{2,\text{det}}, D_L)\left| \frac{\partial(m_{1,\text{det}}, m_{2,\text{det}}, D_L)}{\partial(m_{1,\text{src}}, m_{2,\text{src}}, z)} \right| = D_L^2 (1 + z)^2 \frac{dD_L}{dz}.
\]

With this in mind, we construct a Gaussian kernel density estimate over the source-frame chirp mass and mass ratio $(m_c, q)$ using weights inversely proportional to the transformed prior~\citep{callister2021thesauruscommonpriorsgravitationalwave}:
\[
w_i \propto \frac{1}{\pi_{\text{PE}}(m_{1,i}, m_{2,i}, z_i)}\frac{m_c}{m_1^2}.
\] 
This reweighting yields an approximation to the likelihood function marginalized over all other parameters. 

\section{Full Posterior Distributions}
\label{appendix:full_posterior}
\begin{figure*}
\centering
\begin{overpic}[width=\textwidth]{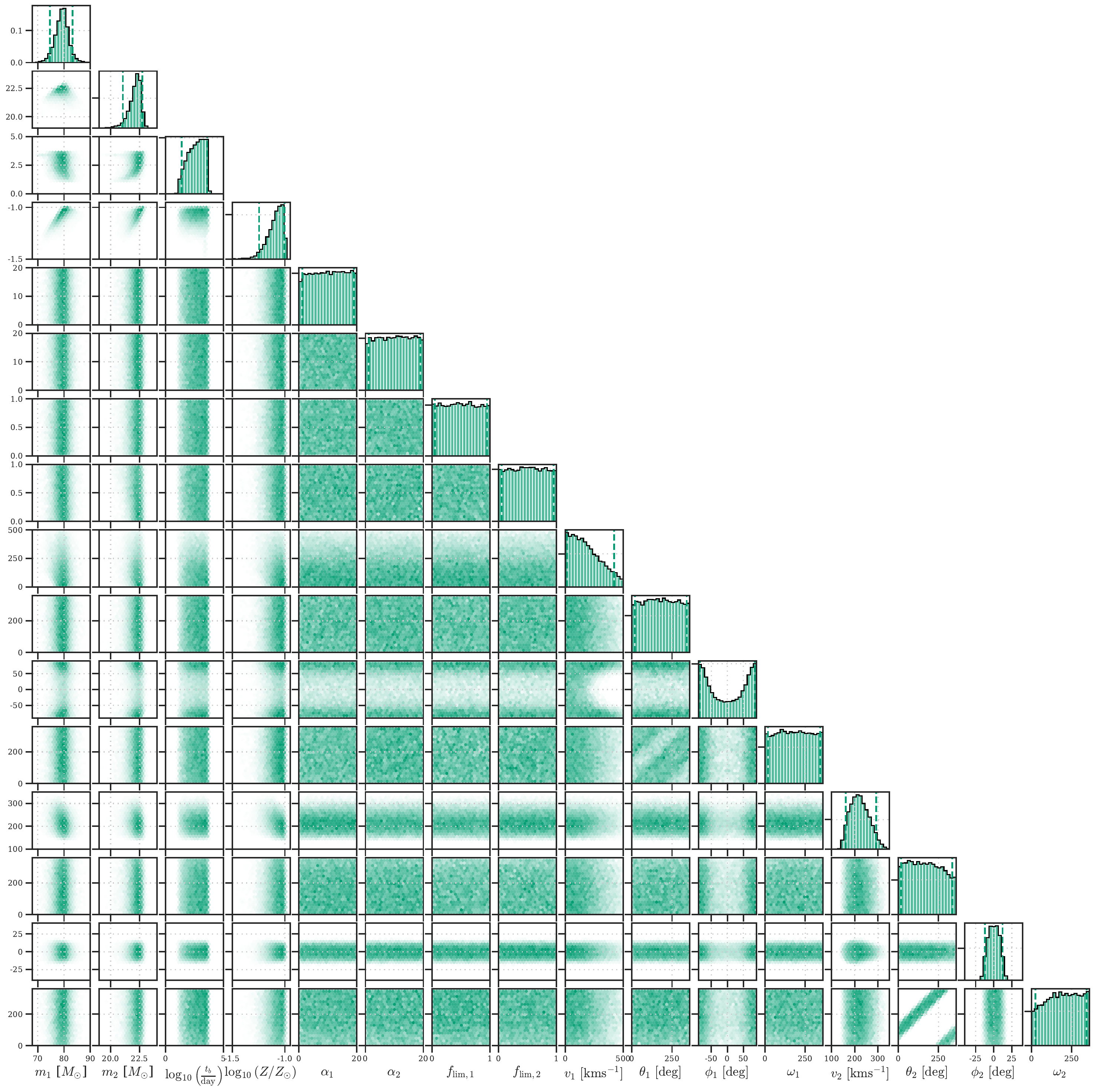}
    % Place inset at (x,y) in percent of main figure width/height
    \put(60,60){ % adjust these numbers to position inset
        \includegraphics[width=0.4\textwidth]{evolved_corner.pdf}
    }
\end{overpic}
\caption{Posterior distribution over the full parameter space defined by our binary population synthesis model. This includes the initial conditions of the binary stars at ZAMS ($m_1$, $m_2$, $\log_{10}t$, $\log_{10}Z$), the astrophysical hyperparameters governing binary interactions ($\alpha_1$, $\alpha_2$, $f_{\rm{lim},1}$, $f_{\rm{lim},2}$), and the supernova natal kick parameters for each compact object ($v_i$, $\theta_i$, $\phi_i$, $\omega_i$ for $i=1,2$). The distributions highlight the regions of parameter space that lead to the successful formation and merger of GW190814-like systems. The top right, shows the chirp mass and mass ratios inferred with our analysis (green) consistent with the LVK reported measurement (gray).}
\label{fig:fat-corner}
\end{figure*}

\begin{figure*}
\centering
\begin{overpic}[width=\textwidth]{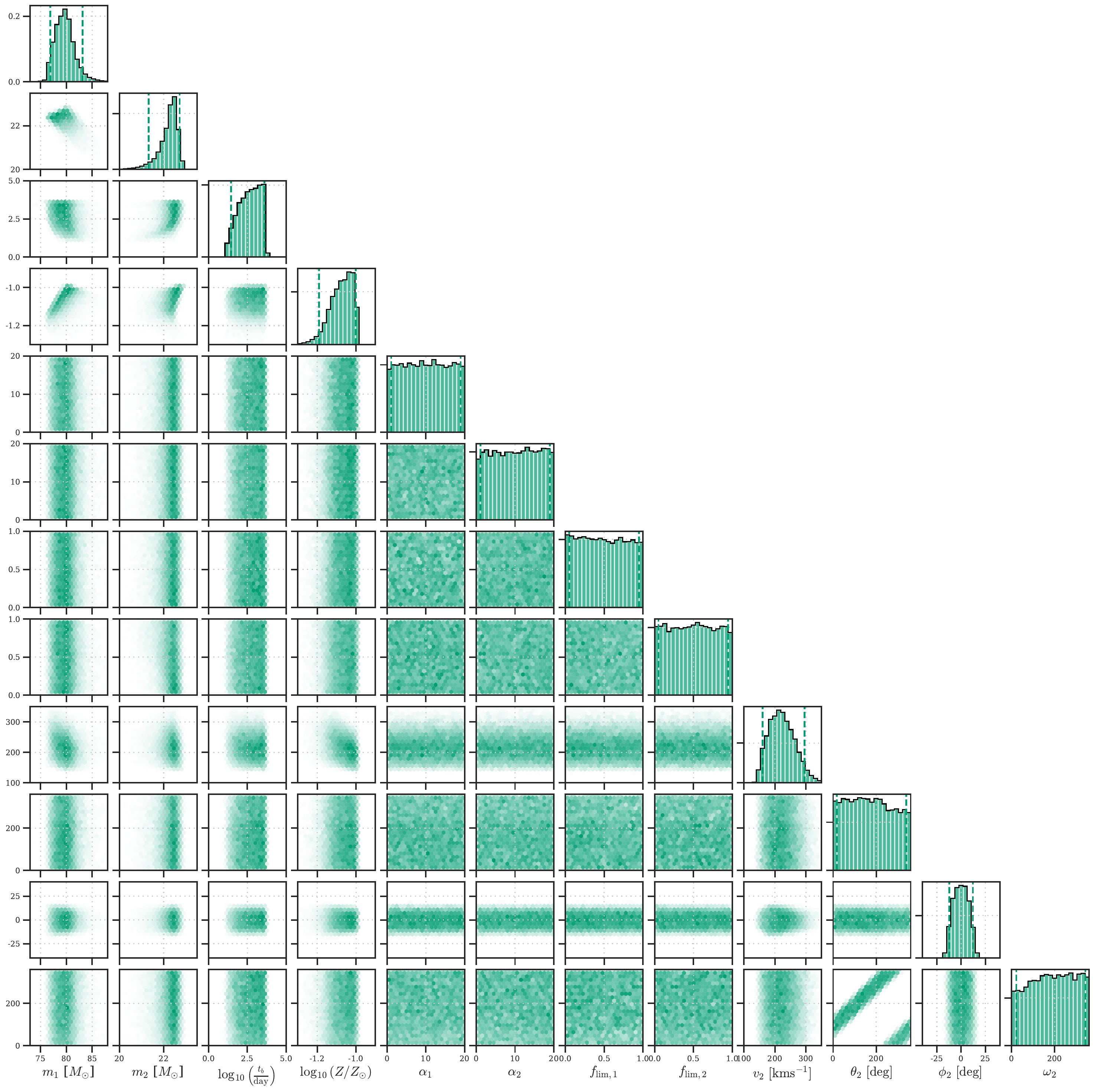}
    % Place inset at (x,y) in percent of main figure width/height
    \put(60,60){ % adjust these numbers to position inset
        \includegraphics[width=0.4\textwidth]{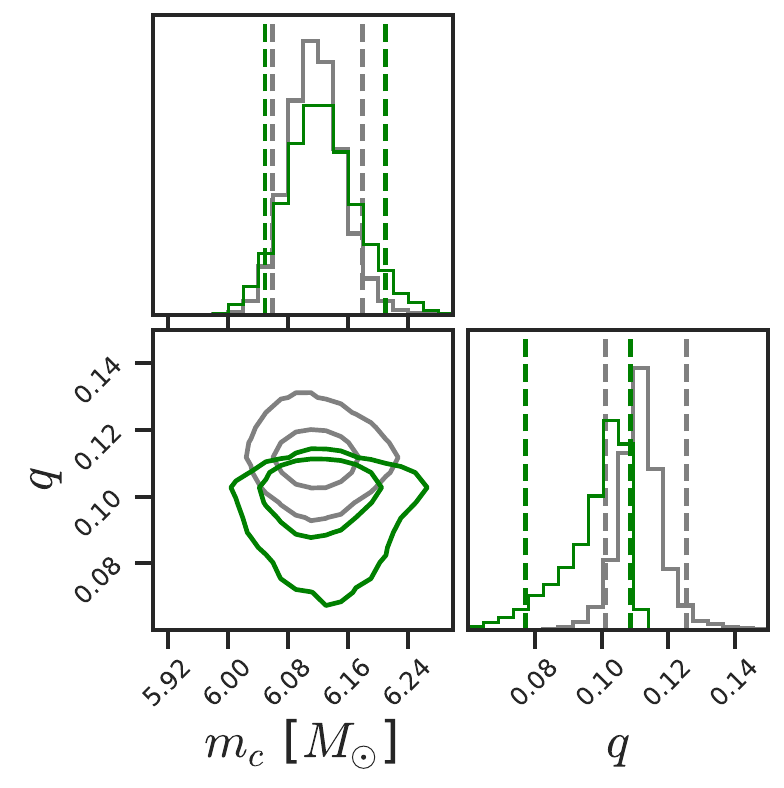}
    }
\end{overpic}
\caption{Posterior distribution over the full parameter space shown in Figure~\ref{fig:fat-corner}, but where we assume that there is not first SN natal kick and so we set $v_{k1} = 0$, $\theta_1 = 0$, $\phi_1 = 0$ and $\omega_1 = 0$. Similarly, the top right posterior shows the inferred BBH masses for GW190814-like systems under this assumption.}
\label{fig:fat-corner-2}
\end{figure*}

In this appendix, we show the full posterior distributions over all model parameters inferred from our simulations of GW190814-like systems for both the full case (shown in Figure~\ref{fig:fat-corner}) and the fixed-kick case (Figure~\ref{fig:fat-corner-2}). In both figures, we show only the $\alpha_1$ and $\alpha_2$ parameters rather than showing the combined $\alpha\lambda_{\rm{CE}}$ parameters since the preferred formation channel does not contain a common envelope evolution and thus $\lambda$ cannot be calculated.  
 
\bibliography{references}{}
\bibliographystyle{aasjournal}

\end{document}